# Closed virial equation-of-state for the hard-disk fluid

Anthony Beris and Leslie V. Woodcock
Department of Chemical Engineering
Colburn Laboratory
University of Delaware,
DE 19716

**A closed virial equation-of-state for the low density fluid phase of hard disks is obtained from the known virial coefficients. The equation exhibits 6-figure accuracy for the thermodynamic (MD) pressure up to the density $\rho\sigma^2 \sim 0.4$. Interpolation of the discrepancy at higher densities indicates a higher-order thermodynamic phase transition at the extensive-intensive free-volume percolation transition previously located by Hoover *et al.* (JCP 70 1837 1979)**

A closed equation-of-state for the virial expansion in powers of density relative to close-packing has been recently proposed for the hard-sphere fluid [1]. This virial expansion is simply written

$$Z = \sum_{n=1}^{\infty} B_n \left(\frac{\rho}{\rho_0}\right)^{(n-1)} \qquad (1)$$

where $Z = pV/Nk_BT$ : $k_B$ is Boltzmann's constant, $B_n$ are the dimensionless coefficients $\rho$ is the number density $(N/V)$, and $\rho_0$ is the crystal close packing density. Equation (1) has been shown to lead to a closed-form equation which is extremely accurate for the hard-sphere fluid; indeed of comparable accuracy to the MD data itself all the way from the ideal gas to the fluid freezing density.

Here we derive a similar equation-of-state for hard disks using only the known virial coefficients up to $B_{10}$. For 2D we use the same nomenclature except now $V$ is the area, and the maximum packing density, $\rho_0 \sigma^2 = 2/\sqrt{3}$ corresponds to the hexagonal close-packed 2D lattice structure. The literature recommended values for all the known virial coefficients for hard disks according to Kolafa and Rottner [2] are reproduced here in TABLE I. In reference [2] the virial coefficients are expressed in an alternative expansion in powers of the packing fraction ( $y \equiv \pi\rho/4$ )

$$Z = \sum_{n=1}^{\infty} \beta_n y^{(n-1)} \qquad (2)$$

where $\beta_n$ is related to $B_n$ in 2D by $B_n = \beta_n (\pi\rho_0/4)^{(n-1)}$.
Also given in TABLE I are the values of the virial coefficients expressed in powers of the density relative to maximum close packing $\rho_0$ in equation (1).

TABLE I    Known virial coefficients of the 2D hard-disk fluid: the values of $\beta_n$ in y-expansion, equation 2, are taken from Kolafa and Rottner [2].

| n | $\beta_n$ [2] | $B_n$ | $B_n - B_{(n-1)}$ | Eq.(3) |
|---|---|---|---|---|
| 1 | 1 | 1 | | |
| 2 | 2 | 1.813799 | 0.813799276 | |
| 3 | 3.128017752 | 2.572691 | 0.758891954 | |
| 4 | 4.257854456 | 3.175912 | 0.603221043 | |
| 5 | 5.33689664 | 3.610153 | 0.43424154 | 0.436659 |
| 6 | 6.3630259 | 3.903550 | 0.293396966 | 0.2902625 |
| 7 | 7.352077 | 4.090396 | 0.18684581 | 0.185693571 |
| 8 | 8.318677 | 4.197288 | 0.106892292 | 0.107266875 |
| 9 | 9.27234 | 4.242903 | 0.045614827 | 0.046268333 |
| 10 | 10.2161 | 4.23953 | -0.00336904 | -0.0025305 |

Additional to the computed virial coefficients, using accurate MD data for disks, over the whole density range up to freezing, Kolafa and Rottner also predicted the higher coefficients $B_{11}$ to $B_{16}$ by fitting to MD data. These fits, however, presuppose that up to the highest density of the MD result fitted, the thermodynamic equation-of-state and the closed virial equation of state are the same. Here, we do not make this assumption in deriving our closed virial equation, but rather, use only the known virial coefficients to derive a closed form. Then, the resulting equation of state can be used to investigate the phase transition whence the thermodynamic pressures from MD first begin to deviate from the virial equation.

Inspection of incremental values of successive virial coefficients, plotted in powers of density relative to crystal close packing as in equation 1; (Fig. 1) shows that beyond ($B_5-B_4$) the increment decrease approximately linearly according to

$$B_n - B_{(n-1)} = \alpha_0 + \alpha/n \quad (n>5) \qquad (3)$$

This interpolation of all the known virial coefficients suggest that since the limiting constant $A_0$ is negative, the virial coefficients will eventually become negative and the corresponding virial equation-of-state will be continuous in all its derivatives, eventually showing a negative pressure, with the first pole at $\rho_o$. Equation (3) with the parameters obtained from Figure 1 predicts the first negative coefficient for D=2 is $B_{31}$.

This observation is now consistent with the known fact that virial coefficients for all hard hyper-spheres of higher dimensionality than 1 eventually go negative. This has been known for almost 50 years since Ree and Hoover calculated $B_4$ for system of various dimension (D) up to D=9 and found $B_4$ to be negative in the two cases of D=8 and D=9. The most recent closed-virial equation for spheres predicts the first negative coefficient in 3D is $B_{19}$. Numerical values for $B_n$ in intermediate D between D= 4 - 7 show that $B_n$ also go negative between $B_4$ and $B_{12}$, and, for some values of D can oscillate in sign. The 10$^{th}$ virial coefficient is negative for all known dimensions greater than D=4 [4]. Equation (3) for D=2 therefore, is consistent with what is generally known about hyper-sphere virial coefficients in all dimensions.

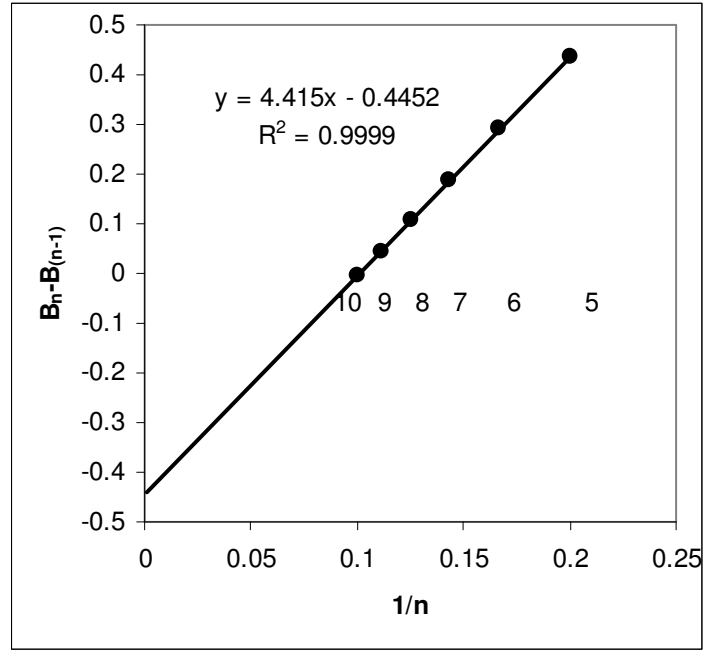

Figure 1: Difference between successive virial coefficients ($B_n$) in the expansion in powers of density relative to close packing: for all known n > 4 the difference $B_n - B_{(n-1)}$ decreases as 1/n, and approaches the constant $\alpha_0 = -0.4452$ when $n \to \infty$.

If we use $\rho^* \equiv (\rho/\rho_0)$ the closed-virial equation-of-state for disks then takes the form (APPENDIX 1)

$$Z = \sum_{n=1}^{m}\left(B_n - \frac{\alpha}{n(1-\rho^*)}\right)\rho^{*(n-1)} + \left(B_m(1-\rho^*)+\alpha_0\right)\frac{\rho^{*m}}{(1-\rho^*)^2} - \frac{\alpha}{\rho^*(1-\rho^*)}\ln(1-\rho^*) \quad (4)$$

where $B_m$ is the highest known virial coefficient, presently $B_{10}$.

Using only the known coefficients $B_5$ to $B_{10}$ of Kolafa and Rottner we obtain the limiting value of $B_n - B_{(n-1)}$, i.e. $\alpha_0 = -0.4415$, and the slope $\alpha = 4.4150$. It is also interesting to note that the constant $\alpha_0$ is very close to half of the close-packed scaled volume $V_0/(2\sigma^2) = \sqrt{3}/4$ Curiously, the corresponding limiting constant for 3D spheres is very close to $V_0/\sigma^3$ [1].

When the pressure is predicted from the closed virial equation (4), can be compared with the thermodynamic pressure from MD computations in the vicinity of melting. In figure 2 we compare with the data of Kolafa and Rottner [2] up to freezing, and our own new MD data, obtained from very long runs of systems on $N=10000$ in the two-phase region and crystal at melting.

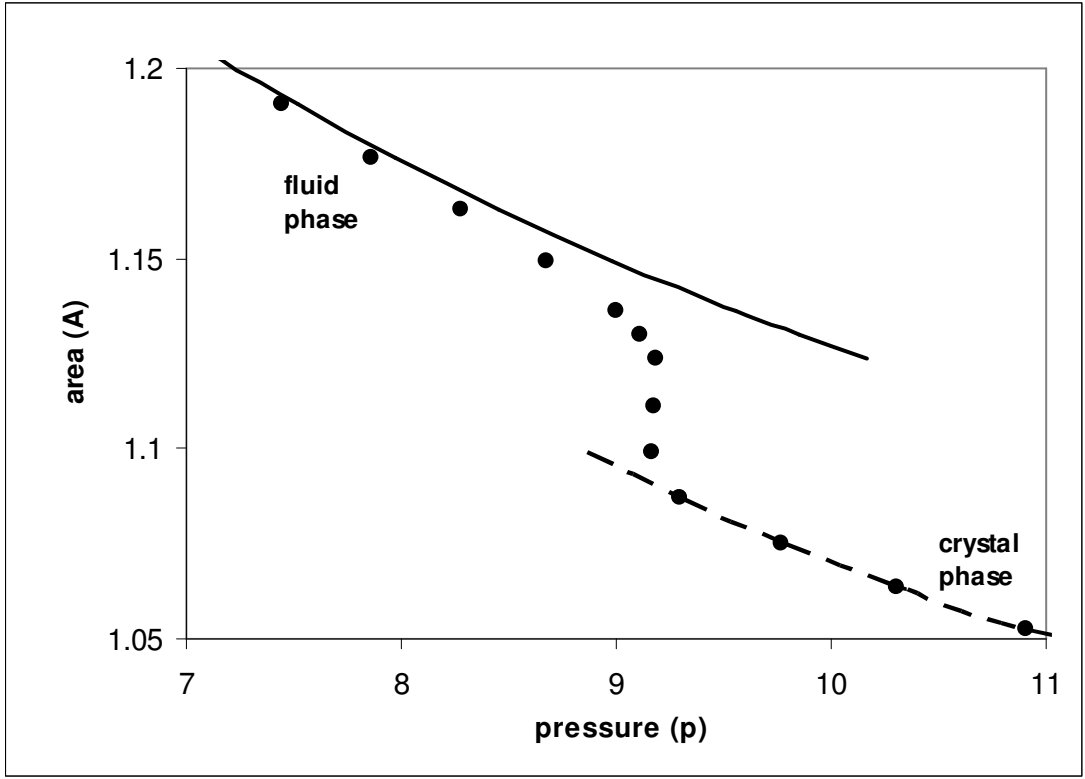

FIG. 2: Equation-of-state for the hard-disk fluid in the vicinity of the first-order freezing transition; virial equation-of-state equation 4 (solid line) crystal equation-of-state [5] (dashed line), MD results Kolafa and Rottner ($p < 9$) and present MD data ($N=10000$: 5000 million collisions each data point) for $p > 9$ and crystal branch.

The comparison between the virial pressure and the fluid thermodynamic pressure in Figure 2 shows that the discrepancy is not simply a "pre-melting phenomena" as has been suggested previously, or even an artifact of MD small systems [7]. A previous closed equation [6] used fitted coefficients $B_{11}$-$B_{15}$ which prejudiced the comparisons.

The present comparison shows that the deviation is real, and is beginning at a much lower density. This begs the question: "where is the first deviation and what is the underlying science"? Since the virial expansion becomes exact at very low density; with convergence up to $\rho\sigma^2 = 0.025$ having been rigorously proven by Lebowitz and Penrose [6], and since the virial equation is continuous in all its derivatives, if the thermodynamic pressure deviates, it must be signaled by a thermodynamic phase transition of second or higher-order.

To investigate this possibility, we have plotted the deviation of equation (4) from MD pressures as a function of density for all the MD data points above $\rho\sigma^2 = 0.4$ in Figure 3. This plot suggests that the deviation is originating in the vicinity of that found by Hoover *et al.* to be the onset of the free volume percolation transition for the hard-disk fluid [8].

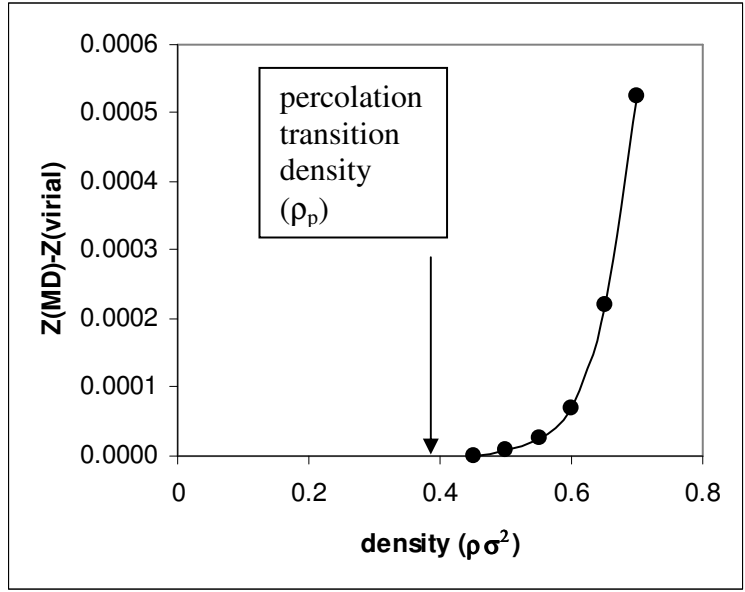

Figure 3: Density dependence of pressure difference between closed-virial equation-of-state (equation (4): *m*=10) and thermodynamic pressures obtained from MD simulations by Kolafa and Rottner (Ref.2).The percolation transition density determined by Hoover et al is indicated by the vertical arrow.

This percolation transition can be defined as the density at which the mean accessible configurational integral of a single disk, within the equilibrium ensemble, changes from being extensive at low density to intensive at high density. This free volume ($v_f$) is related to the thermodynamic pressure according to

$$Z = 1 + \frac{\sigma}{2D}\left\langle\frac{s_f}{v_f}\right\rangle \qquad (5)$$

and $s_f$ is the surface area of the hole in which the disk is trapped at high density, or the networked union of holes at densities below the percolation transition. The density at which $<v_f>$ changes from being an intensive state function to an extensive state function defines the percolation transition. It follows from equation (5) that $<s_f>$ must change likewise at the same transition point.

At low density, the cell-free volume ($<v_f>$) becomes equivalent to the so-called spare volume ($<v_s>$) , which is the relative probability of adding an extra sphere anywhere in an equilibrium configuration, and hence is exactly related to the excess Gibbs chemical potential ($G$), for large $N$

$$\mu \equiv \frac{dG}{dN} = -k_B T \ln\left(\frac{\langle v_s \rangle}{V}\right) \qquad (6)$$

Now, we deduce that if equations (5) and (6) are rigorous, and for densities below than $\rho_p$ (the percolation transition)

$$\langle v_f \rangle = \langle v_s \rangle; \quad \rho < \rho_p \qquad (7a)$$

whereas for densities above the percolation transition we have

$$\langle v_f \rangle = \mathcal{O}\left(\langle v_s \rangle / N\right); \quad \rho > \rho_p \qquad (7b)$$

It seems that there is a discontinuity in a higher derivative of $<v_s>$ in equation (6) and a deviation between thermodynamic pressure and the cluster expansion at the point when a cluster of size $N$ spans the whole system. Thermodynamically, fluctuations in number density are exactly related to the second-derivative of the chemical potential $d\mu/d\rho$. At the percolation transition density when one cluster spans the whole system, a class of density fluctuations become frozen out. One therefore expects not a second, but a third-order thermodynamic transition, i.e. a discontinuity in the derivative of the isothermal compressibility, which is the equivalent of heat capacity $Cp$ for disks as a consequence of simple scaling. This observation may explain the apparent high accuracy of the closed-virial equation in 3D almost up to freezing, even though it may belong to a different phase at liquid densities. The percolation transition is weaker in three dimensions.

We should now have a closer look at the hard-sphere percolation transition in 3D as this result has further implications for the general development of theories of the liquid state and the origin of critical point phenomena. Traditional approaches based upon integral equation closures for hard spheres all assume the essential correctness of the Mayer cluster expansion with continuity up to liquid densities. It now appears that this may not be so.

**APPENDIX**: Derivation of closed virial equation for hard-disk fluid:

If $x = \rho/\rho_0$, from equations (1) and (3) we have

$$Z = \sum_{n=1}^{m} B_n x^{(n-1)} + \sum_{n=m+1}^{\infty} \left( \sum_{l=m+1}^{n} \left( \alpha_0 + \frac{\alpha}{l} \right) + B_m \right) x^{(n-1)} \tag{A1}$$

However, note that we can use the following expressions for the summation of the geometric series appearing in the second term in the above equation

$$\begin{aligned}
\sum_{n=m+1}^{\infty} \left( \sum_{l=m+1}^{n} (\alpha_0) \right) x^{(n-1)} &= \sum_{n=m+1}^{\infty} \left( [n-m]\alpha_0 \right) x^{(n-1)} \\
&= \alpha_0 x^m \sum_{n=m+1}^{\infty} \left( [n-m] \right) x^{(n-m-1)} \\
&= \alpha_0 x^m \sum_{k=0}^{\infty} k x^{(k-1)} = \alpha_0 x^m \frac{d}{dx} \left( \sum_{k=0}^{\infty} x^k \right) \\
&= \alpha_0 x^m \frac{d}{dx} \left( \frac{1}{1-x} \right) = \frac{\alpha_0 x^m}{(1-x)^2}
\end{aligned} \tag{A2}$$

$$\begin{aligned}
\sum_{n=m+1}^{\infty} \left( \sum_{l=m+1}^{n} \left( \frac{\alpha}{l} \right) \right) x^{(n-1)} &= \alpha \sum_{l=m+1}^{\infty} \left( \frac{1}{l} \sum_{n=l}^{\infty} x^{(n-1)} \right) = \alpha \sum_{l=m+1}^{\infty} \left( \frac{1}{l} \frac{x^{(l-1)}}{(1-x)} \right) \\
&= \frac{\alpha}{(1-x)} \left[ \frac{1}{x} \sum_{l=1}^{\infty} \left( \frac{1}{l} x^l \right) - \sum_{l=1}^{m} \left( \frac{1}{l} x^{(l-1)} \right) \right] \\
&= \frac{\alpha}{(1-x)} \left[ \frac{1}{x} \int_0^x \left( \sum_{l=1}^{\infty} z^{(l-1)} \right) dz - \sum_{l=1}^{m} \left( \frac{1}{l} x^{(l-1)} \right) \right] \\
&= \frac{\alpha}{(1-x)} \left[ \frac{1}{x} \int_0^x \left( \frac{1}{1-z} \right) dz - \sum_{l=1}^{m} \left( \frac{1}{l} x^{(l-1)} \right) \right] \\
&= \frac{\alpha}{(1-x)} \left[ -\frac{\ln(1-x)}{x} - \sum_{l=1}^{m} \left( \frac{1}{l} x^{(l-1)} \right) \right]
\end{aligned} \tag{A3}$$

$$\sum_{n=m+1}^{\infty} (B_m) x^{(n-1)} = B_m \sum_{n=m+1}^{\infty} x^{(n-1)} = B_m \frac{x^m}{1-x} \tag{A4}$$

When Eqs. (A2-A4) are used within Eq. (A1) it is easy to see that we get Eq. (4).